\documentclass[conference]{IEEEtran}
\IEEEoverridecommandlockouts
\usepackage{cite}
\usepackage{amsmath,amssymb,amsfonts}
\usepackage{algorithmic}
\usepackage{graphicx}
\usepackage[hyphens]{url}
\usepackage{textcomp}
\usepackage{xcolor}
\usepackage{lipsum}
\usepackage[caption=false]{subfig}
\def\BibTeX{{\rm B\kern-.05em{\sc i\kern-.025em b}\kern-.08em
    T\kern-.1667em\lower.7ex\hbox{E}\kern-.125emX}}
\begin{document}
\begin{sloppypar}

\title{IoT Wallet: Machine Learning-based Sensor Portfolio Application\\
}

\author{
    \IEEEauthorblockN{Petar Šolić\IEEEauthorrefmark{1}, Ante Lojić Kapetanović\IEEEauthorrefmark{1}, Tomislav Županović\IEEEauthorrefmark{1}, Ivo Kovačević\IEEEauthorrefmark{1}, Toni Perković\IEEEauthorrefmark{1}, Petar Popovski\IEEEauthorrefmark{2}}
    \IEEEauthorblockA{\IEEEauthorrefmark{1}\textit{Faculty of Electrical Engineering, Mechanical Engineering and Naval Architecture}\\
    \textit{University of Split}\\
    \big\{psolic, alojic00, tzupan01, ikovac01, toperkov\big\}@fesb.hr}
    \IEEEauthorblockA{\IEEEauthorrefmark{2}\textit{Department of Electronic Systems}\\
    \textit{Aalborg University}\\
    petarp@aau.es.dk}
}

\maketitle
\begin{abstract}
In this paper an application for building sensor wallet is presented. Currently, given system collects sensor data from The Things Network (TTN) cloud system, stores the data into the Influx database and presents the processed data to the user dashboard. Based on the type of the user, data can be viewed-only, controlled or the top user can register the sensor to the system. Moreover, the system can notify users based on the rules that can be adjusted through the user interface. The special feature of the system is the machine learning service that can be used in various scenarios and is presented throughout the case study that gives a novel approach to estimate soil moisture from the signal strength of a given underground LoRa beacon node.
\end{abstract}

\begin{IEEEkeywords}
Internet of Things, LoRa, deep learning, time series modeling, long short-term memory neural networks
\end{IEEEkeywords}

\section{Introduction}
\label{sec.introduction}

Internet of Things (IoT) represents a paradigm in the world of interconnected devices in which sensors, actuators, smartphones communicate with each other using appropriate wireless technologies. In such an environment, users can receive feedback about the physical world that surrounds them, allowing the interaction with it and exchange such data with the digital world~\cite{Mahdavinejad18MLI}. IoT applications are used today in various aspects of the industry, such as wireless sensor networks, data mining, assisted living, etc. giving rise to the concept of Smart City. Application of IoT can be found in every aspect of everyday lives, ranging from the fields of automation, industrial manufacturing, logistics, business/process management, intelligent transportation of people and goods or environmental monitoring~\cite{Atzori10IoTS, Barile18Real-Time}.

IoT denotes a concept in which a large number of interconnected devices create novel applications. Despite the industry forecasts of 50 billion interconnected devices, only 9 billion IoT devices have been materialized until 2020\footnote{\url{https://www.vanillaplus.com/2020/01/16/50049-missing-41bn-iot-devices-biggest-prediction-miss-history/}}. For the IoT to come to life in full, it is necessary to bring the functionality closer to the end-user through a flexible platform where information on the status of "things" being monitored can be obtained and the user is informed if changes occur. Likewise, flexibility must be manifested through the integration with future services where Machine Learning (ML) in IoT applications allows for different estimations and predictions based on data coming from sensor devices.

For example, IoT applications are especially suited for living environments such as agricultural, where the irrigation plays an extremely important role~\cite{Balducci18MLA}. Existing solutions include the use of battery-operated sensing devices, while sensor data is transmitted using the appropriate wireless technology. Recent advancements in Low Power Wide Area Network (LPWAN) gave rise to radio technologies, e.g. LoRa (LoRaWAN), NB-IoT and Sigfox, that are suitable for sporadic transmissions of small sensor data packets over large distances, making them ideal for livestock farming, flood monitoring and/or smart irrigation systems~\cite{Brewster17IoT, Perkovic20Meeting}. In LPWAN architecture application-specific end nodes reach their network server via a gateway. From there, the data are routed to the respective application server. Such architecture gave rise to the commercial LPWAN service providers, e.g. The Things Network (TTN)\footnote{\url{https://www.thethingsnetwork.org/}} and LORIOT\footnote{\url{https://www.loriot.io/}}, that employ functionalities of LoRaWAN and/or application server. Consequently, application specific platforms that allow a user to visualize and possibly control sensor device status have been developed. For example, myDevices Cayenne\footnote{\url{https://www.thethingsnetwork.org/docs/applications/cayenne/}} and Ubidots\footnote{\url{https://ubidots.com/}} provide a service to visualize real-time and historical data sent over The Things Network, such as temperature monitoring, occupancy, predictive maintenance, etc. Libelium, on the other hand, has developed a cloud platform for monitoring a plethora of sensing devices provided by Libelium company\footnote{\url{https://cloud.libelium.com/}}.

All these systems as introduced previously enable users the possibility to insert new proprietary sensor devices. Similarly, as in systems described above, the proposed system allows users to collect the sensor data from TTN cloud and store it in a time-series Influx database. Such a feature is given by default in the majority of services as it allows users to visualize the data using appropriate dashboards. In addition to related services, the proposed solution allows the user to send commands to the sensor data, such as the wake-up period, time synchronization, etc. A special feature of the collected data is a ML service that realizes a novel approach in the big data analysis, allowing users to work on various case studies aimed at solving the domain specific problems. Hence, in this paper an IoT Wallet is introduced, whereas a case study is the soil moisture prediction from a signal strength of an underground LoRa beacon. Using the double prediction of the feed-forward neural network and the long short-term memory (LSTM) network, highly accurate prediction of the soil moisture is achieved solely based on the publicly available data acquired from the State Hydrometeorological Institute and characteristics of the signal received from LoRa end devices.

\section{LoRa-based soil moisture Sensor}
\label{sec:lora}

Using signal strength measurements from a beacon device with the sensor data from other devices, such as air humidity, temperature, and pressure, soil moisture is estimated using related ML techniques\cite{Dujic20MLS}. Hence, soil moisture sensor devices could be replaced with a simple underground LoRa beacon end-device. In this paper, LoRa technology is employed for transmitting communication data from sensor devices to the base station (e.g. soil moisture, air temperature, and humidity). LoRa as a representative of LPWAN allows battery-operated sensors to communicate low throughput data over long distances, making it suitable for applications in scenarios such as agriculture monitoring~\cite{LiedmannHW18}. %In the architecture of LoRaWAN protocol, end-devices are connected to one or many gateways which are in turn directly connected to a network and application server. Hence, the gateway simply acts as a relay device which allows the network server to further forward the information to corresponding application servers for processing.

\subsection{Data Collection from LoRa-based sensor device}\label{sec:implementation}

Namely, soil moisture, along with signal strength measurements (RSSI and SNR) were collected from humidity sensor device that was buried 14 cm below the ground level. %Although in final version underground LoRa beacon device will be employed (without soil moisture sensor), here soil moisture sensor was employed to collect ground truth information about soil moisture, which will be compared against estimated soil moisture values using related machine learning techniques.
Besides, air humidity and temperature were collected from another LoRa-based sensor device that was placed 3 m over the ground in constant shade. The core of both sensor devices is Arduino Pro Mini (ATmega328P) that operates at 3.5V. For LoRaWAN communication, an RFM95W module that uses SX1276 chip was used that operates at 868 MHz. In our implementation spring antenna was vertically oriented with $+14$dB transmission power. Also, TPL5110 module was used to preserve energy during inactive periods, where the module simply cuts off power during the inactive period, reducing the overall consumption of the sensor device. The TPL5110 timer was set to power up Arduino every 10 minutes. For the measurement of soil moisture, an I2C soil moisture sensor was employed with capacitive sensing\footnote{\url{https://www.whiteboxes.ch/shop/i2c-soil-moisture-sensor/}}, while for the measurement of air humidity and temperature, SHT-10 mesh protected sensor device was used. Besides, publicly available data was such as air pressures were acquired from the State Hydrometeorological Institute for the city where our system is deployed.

\begin{figure}[t]
\centering
   \includegraphics[width=0.75\linewidth]{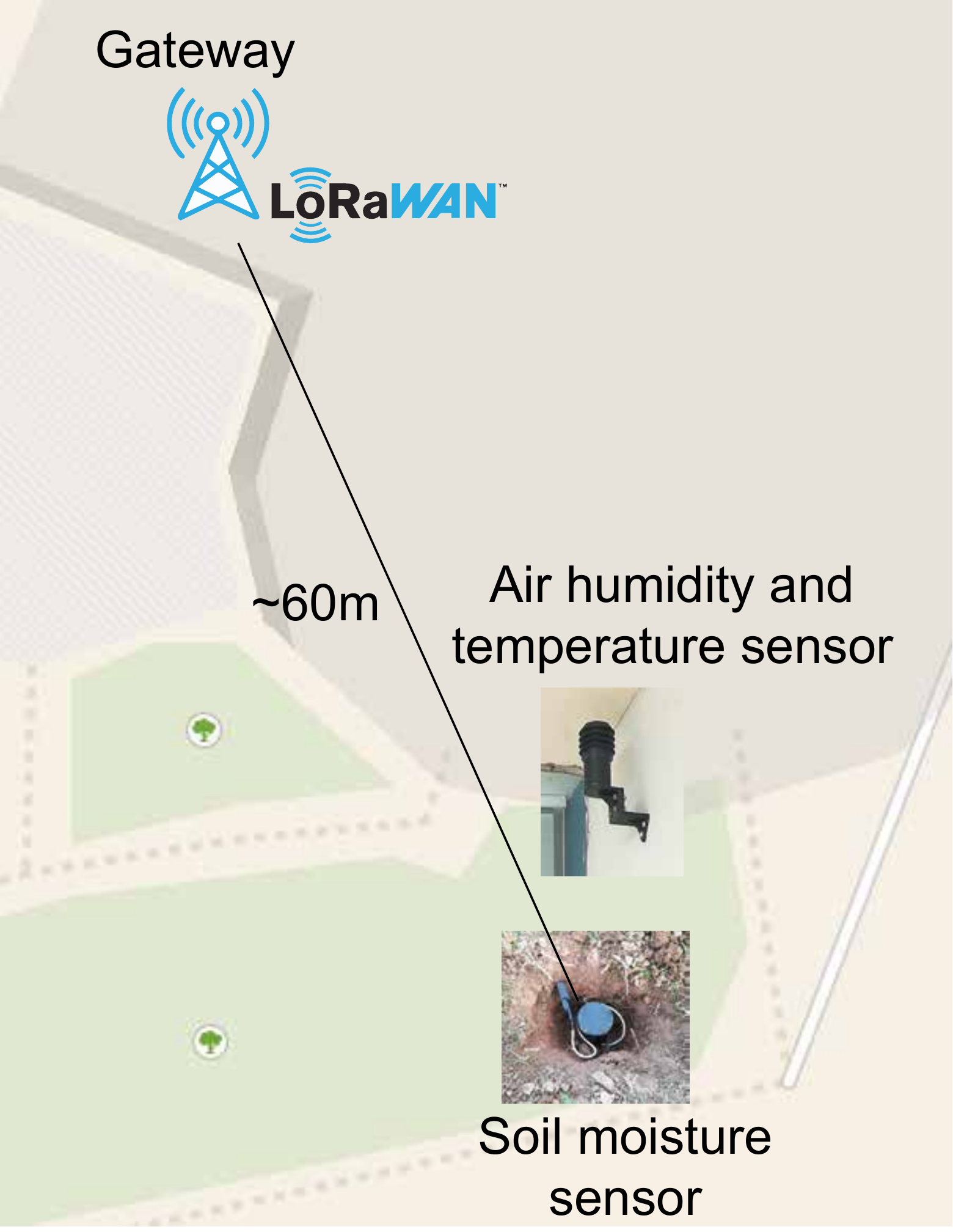}
   \caption{Implementation of LoRa-based sensor devices.}
   \label{fig:implementation}
\end{figure}

As a LoRaWAN gateway, an indoor Raspberry Pi-based gateway device was employed that forwards messages to The Things Network (TTN) cloud infrastructure. Our gateway uses RAK831 concentrator with Procom CXL 900-6LW-NB, 8 dBi gain, 868 MHz, vertically polarized, omnidirectional antenna placed around 60 meters from soil moisture sensor device (Figure~\ref{fig:implementation}).

\begin{figure}[t]
\centering
   \includegraphics[width=0.73\linewidth]{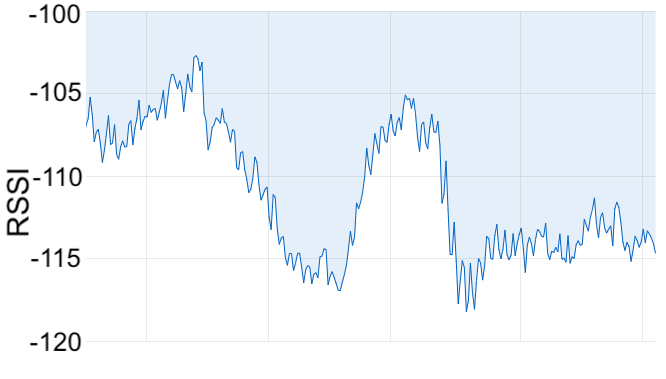}
   \includegraphics[width=0.73\linewidth]{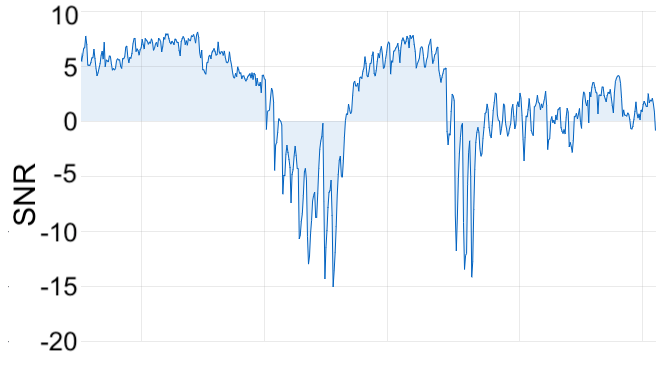}
   \includegraphics[width=0.73\linewidth]{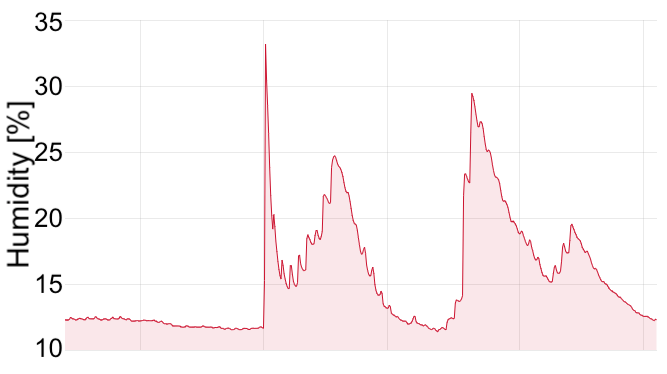}
   \includegraphics[width=0.73\linewidth]{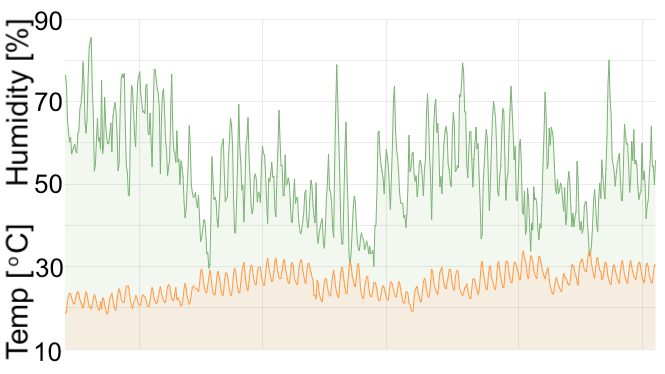}
   \includegraphics[width=0.73\linewidth]{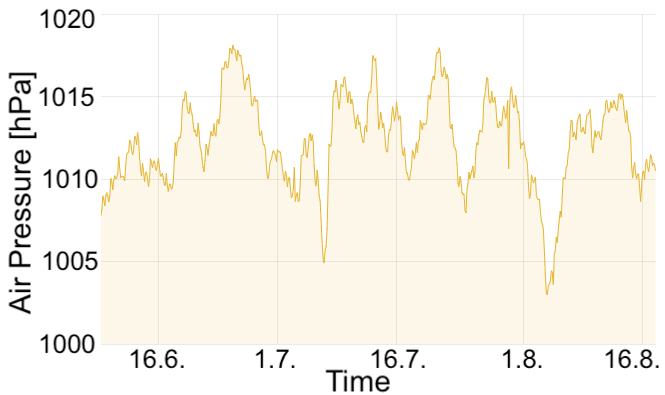}
   \caption{Snapshot of RSSI and SNR signal captured on LoRaWAN gateway from soil moisture sensor, along with I2C sensor measures of soil moisture. Another LoRaWAN sensor sends information about air humidity and temperature, while State Hydrometeorological Institute gives information regarding air pressure.}
   \label{fig:measurements}
\end{figure}

Once the message arrives at the gateway, it is forwarded to the TTN Network and Application server. Furthermore, TTN allows message forwarding from their infrastructure to our dedicated IoT Wallet using MQTT protocol, which is described more in detail in Section~\ref{sec.softarch}. Figure~\ref{fig:measurements} shows a snapshot of soil moisture along with RSSI and SNR values captured on the gateway. Besides, air humidity and temperature from the second LoRa-based sensor devices are collected, along with air pressure data collected from State Hydrometeorological Institute. As can be seen, when the soil moisture increases, both RSSI and SNR signal values drop, showing the tight bound between these values. In addition, in order to improve the prediction of soil moisture, information on air temperature, humidity and pressure are collected as well.

\begin{figure*}[htbp]
   \includegraphics[width=\textwidth]{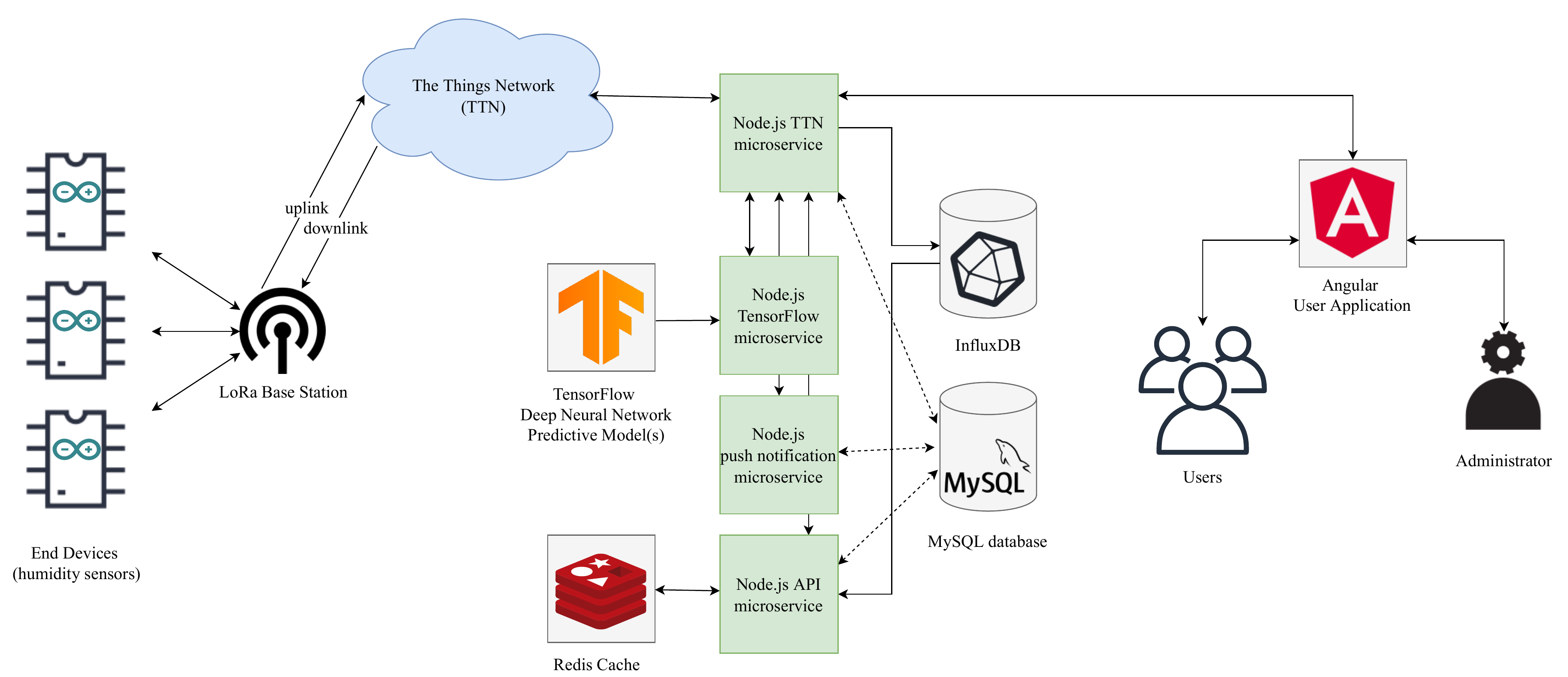}
   \caption{High-level overview of the current implementation of the solution.}
   \label{fig.system}
\end{figure*}

\section{IoT Wallet Architecture - Overview and Functionalities}
\label{sec.softarch}

The Things Network (TTN) microservice is set up to listen for the LoRa sensor data using MQTT protocol, as depicted in Figure~\ref{fig.system}. Once the signal arrives, the TTN microservice captures and stores the LoRa uplink data into the Influx database. Also, the TTN microservice is capable of listening for LoRa-based downlink messages, which are redirected from the client application to the TTN cloud. The purpose of downlink is to control sensor’s behavior within the application itself. 

Immediately after the data are stored in the Influx database, TTN microservice sends sensor data to the ML-based forecasting microservice. The data are processed through ML model which forecast the soil moisture values for the following moment or the sequence of following moments, more details are presented in the Section \ref{sec.ml}. 

The push notification microservice sends notifications to the user. Notifications are triggered if sensor value is greater than, less than or equal to the value that the user has specified. There is also an option for cumulative notifications for the specific time period. In this case, the user enters a cumulative period, the operator and the value which will trigger notification.

Lastly, the API microservice manages the complete logic of the application. The API microservice has GraphQL server running and listening for all traffic on the pre-defined port. The client application communicates with the API microservice and serves the data to the end user. Main API endpoints are the user, sensors, the sensor type, downlink control etc. Client application is built within the Angular framework. Ionic framework is set up as a top layer of the Angular application and it enables the cross platform deployment. This opens the possibility of serving the application for the web, Android and iOS, simultaneously.

\section{Machine Learning in IoT (Wallet)}
\label{sec.ml}
Data mining, big data analysis and ML are paradigms of particular importance for the current state of IoT and the Industrial IoT.
From smart traffic, health, environment and agriculture all to control, security and forecasting of different natural phenomena, these services can be enhanced and optimized by analyzing the smart data generated and collected in their respective domains \cite{Mahdavinejad_ML_2018}. 
The case-study presented in the following subsection is the soil moisture prediction from the signal strength indicator.

\subsection{Case-study: soil moisture Prediction from Signal Strength}
\label{sec.casestudy}
In this work, the idea is to replace expensive capacitive-based soil moisture sensors, most often unreliable and in need of human intervention in terms of calibration and battery change, with simple LoRa beacon end-devices. LoRa beacon lights up sporadically and sends a short signal to the base station where channel information is recorded and the soil moisture is predicted, assuming a nonlinear dependence of the received signal strength by means of RSSI and SNR and the soil moisture.
We incorporate learning by two approaches: training a simple yet powerful deep neural network to extract the current knowledge from the data, and training a LSTM network which allows the historical, long-term dependencies to be captured. 

\begin{figure}[ht]
\centering
   \includegraphics[width=\linewidth]{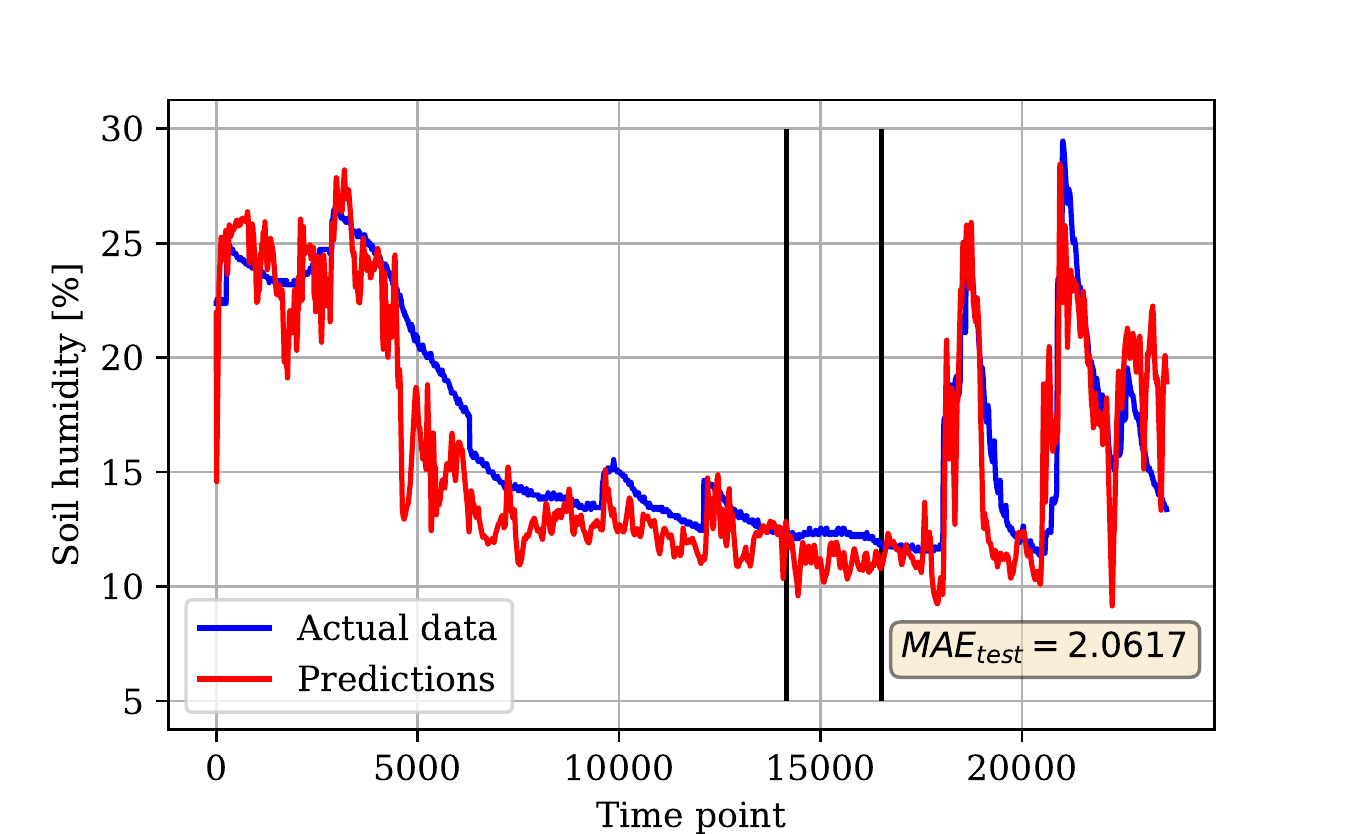}
   \caption{The actual data are shown in blue, while the predicted data using a feed-forward neural network are shown in red on the training set (curves to the left of the first black vertical line), validation set (curves between the two black vertical lines) and test set (curves to the right of the second black vertical line), respectively.}
   \label{fig.nn}
\end{figure}

In the first approach, a deep feed-forward neural network with $2$ hidden layers, containing $128$ and $64$ units, respectively, is trained on $23592$ data points. The data have been collected, captured and stored on every $10^{th}$ minute from the mid February 2020 to mid August 2020. Each input consisted of $5$ features, which may be separated into the two logical sets. The first set is closely reflecting the nature of the signal (RSSI and SNR) and the second set is describing current atmospheric conditions (air temperature, air humidity and air pressure) of the area where the LoRa sensors are located. The neural network is trained for $500$ epochs using the batch size of 32. Optimization of the Mean Square Logarithmic Error (MSLE) loss function is performed using ADAM optimization technique introduced in \cite{kingma_adam_2014} with learning rate initially set to $0.0001$. The weight matrix initialization is performed using a well-known Xavier initialization scheme \cite{glorot_bengio_2010}. Each unit is activated using Exponential Linear Unit (ELU) activation function. Before the training itself, the data are split into the training, validation and test set. Additional pre-processing of the data include double normalization. The first step is normalizing the training data by subtracting the mean value. This provides a more general approach to learning and opens the door for applying the same (or minimally re-trained) model to other LoRa sensors in significantly different conditions. The second step is scaling the training data between 0 and 1 so that the order of magnitude of the values that the model is fed with and the weight of the model itself are the same. This allows avoiding the issue of the exploding gradients and creates stable and secure training conditions. In Figure \ref{fig.nn}, the inference on the test set is shown where the mean absolute error (MAE) is $2.0617$. 

\begin{figure}[ht]
\centering
   \includegraphics[width=\linewidth]{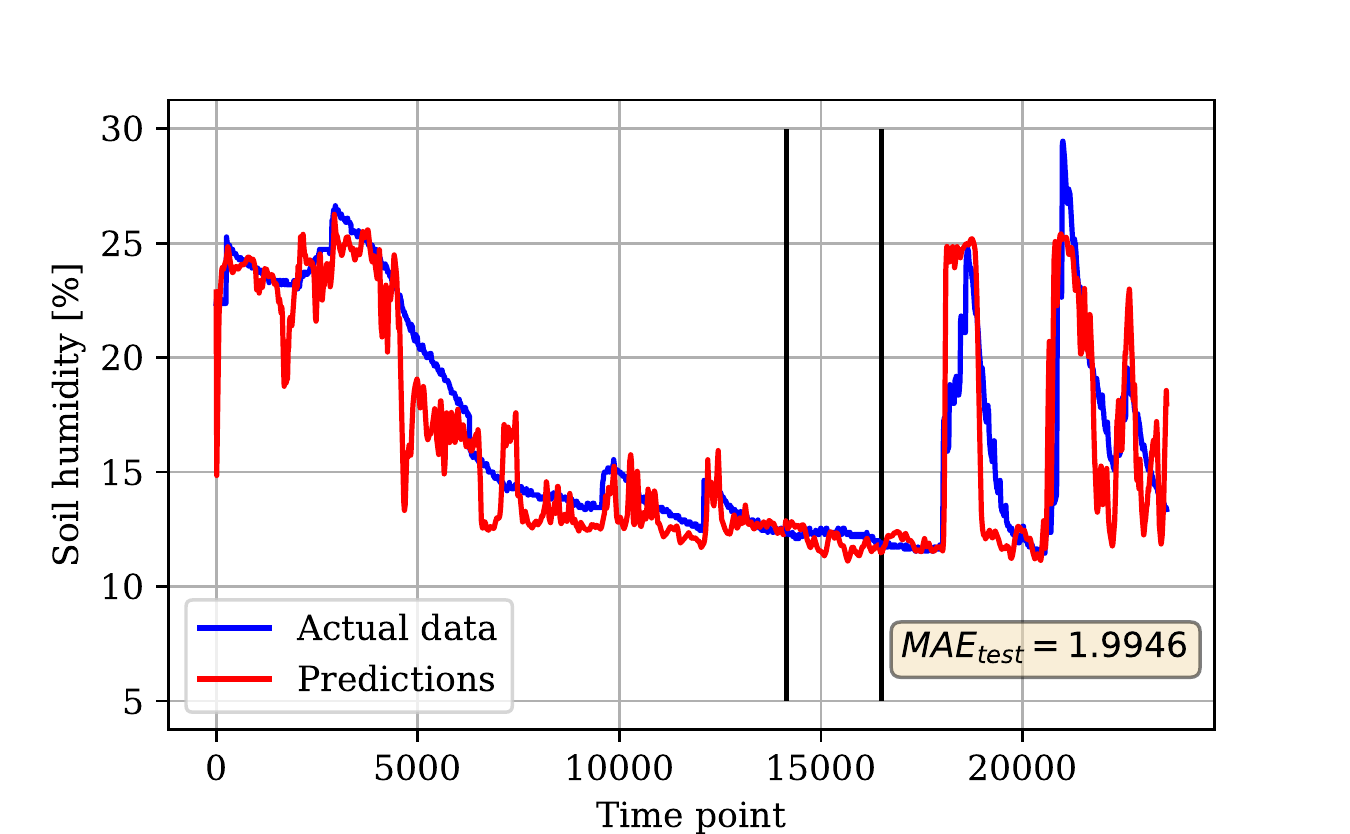}
   \caption{The actual data are shown in blue, while the predicted data using the LSTM network are shown in red on the training set (curves to the left of the first black vertical line), validation set (curves between the two black vertical lines) and test set (curves to the right of the second black vertical line), respectively.}
   \label{fig.lstm}
\end{figure}

In order to capture long-term dependencies in the data, the LSTM network is built and trained \footnote{The first version of the LSTM network is introduced in \cite{Hochreiter_lstm_1997} but since then is upgraded and adjusted to the modern software capabilities \cite{olah_lstm_2015}, which exploit the full potential of the automatic differentiation.}. Because of its recurrent nature, the LSTM network and its units are perfectly adapted to the time-series data (a series of data points indexed in time order and stored at successive equally spaced points in time). The LSTM unit architecture curates the long-term dependency problem with its three-gate configuration allowing information to persist and capturing underlying non-linear pattern in the data, otherwise imperceptible to \textit{vanilla} neural networks. Here, the network consists of a single LSTM layer and a single dense layer with 12 and 20 units, respectively. Training is performed throughout $500$ epochs using the same data and the same input features as for the previous neural network, but here the last $6$ steps are considered in each training step. Scaling and normalization also stayed the same. All hyper-parameters but the activation function stayed the same. The activation function is set by default to Hyperbolic Tangent (\textit{tanh}) function using the LSTM layer deployed in NVIDIA CUDA Deep Neural Network library (cuDNN) adapted for TensorFlow framework \cite{tensorflow_2015}. In Figure \ref{fig.lstm}, the inference on the test set is shown where the MAE is $1.9946$.

\begin{figure}[ht]
\centering
   \includegraphics[width=\linewidth]{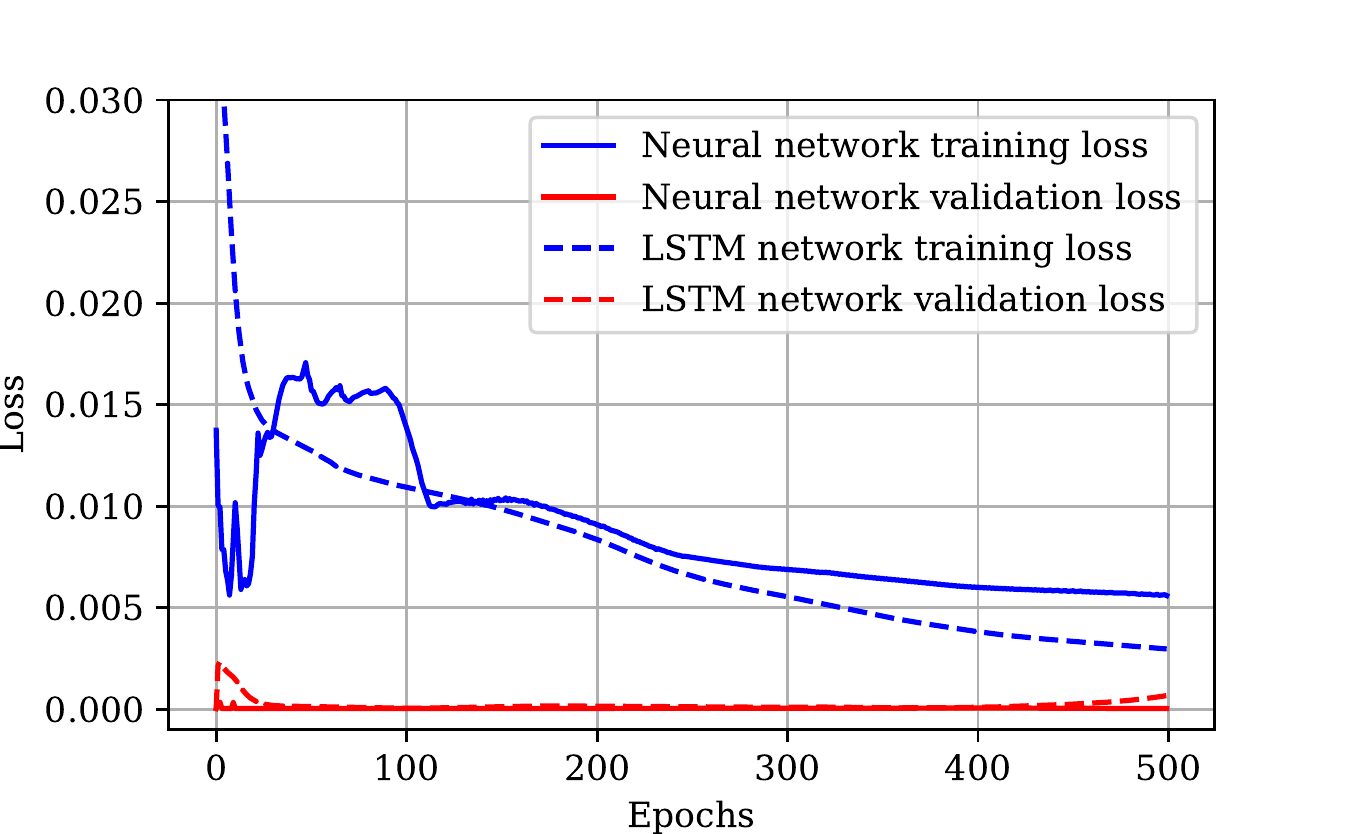}
   \caption{Learning curves by means of the loss change over training epochs of the feed-forward neural network and the LSTM network are shown using blue and red colored full line and blue and red colored dashed line, respectively.}
   \label{fig.loss}
\end{figure}

Learning curves for both previously outlined models - the feed-forward neural network and the LSTM network, are shown in Figure \ref{fig.loss}. A slight overfitting in the case of the LSTM network is visible on the validation curve by the end of the training. However, due to the fact that the overall validation loss is lower than the training loss, overfitting can be overlooked. This phenomenon may happen usually when the training data is more difficult to train on and learn patterns on, while less change occurs in the validation set.

Using the double prediction of the feed-forward neural network and the LSTM network, respectively, in the form of an ensemble of learning, we are able to achieve reliable and accurate prediction of the soil moisture solely on the basis of the publicly available data acquired from the State Hydrometeorological Institute and characteristics of the signal received from LoRa end devices.

\section{Conclusion}
\label{sec.conclusion}

In this paper architecture of IoT Wallet system is introduced. It allows users to manually register the sensor device to TTN service, as well as to visualize the data stored in the database. Besides, depending on the user type, it is allowed to control the sensors, and communicate data towards them. A special feature of the proposed IoT Wallet is a ML feature that allows users to perform predictions from collected sensed data. As a proof-of-concept, soil moisture is predicted from signal strength from an underground LoRa beacon device. As shown in this paper, using the feed-forward neural network and the LSTM network, an accurate prediction of the soil moisture can be estimated based on the signal characteristics of LoRa device along with collected publicly available data from overground air humidity and temperature device along with air pressure collected from Hydrometeorological Institute.

\section*{Acknowledgment}
This Technology Transfer Experiment has received funding from the European Union’s Horizon 2020 research and innovation programme under the TETRAMAX grant agreement no 761349.

\bibliographystyle{IEEEtran}{}
\bibliography{./bibl}

\begin{thebibliography}{10}
\providecommand{\url}[1]{#1}
\csname url@samestyle\endcsname
\providecommand{\newblock}{\relax}
\providecommand{\bibinfo}[2]{#2}
\providecommand{\BIBentrySTDinterwordspacing}{\spaceskip=0pt\relax}
\providecommand{\BIBentryALTinterwordstretchfactor}{4}
\providecommand{\BIBentryALTinterwordspacing}{\spaceskip=\fontdimen2\font plus
\BIBentryALTinterwordstretchfactor\fontdimen3\font minus
  \fontdimen4\font\relax}
\providecommand{\BIBforeignlanguage}[2]{{%
\expandafter\ifx\csname l@#1\endcsname\relax
\typeout{** WARNING: IEEEtran.bst: No hyphenation pattern has been}%
\typeout{** loaded for the language `#1'. Using the pattern for}%
\typeout{** the default language instead.}%
\else
\language=\csname l@#1\endcsname
\fi
#2}}
\providecommand{\BIBdecl}{\relax}
\BIBdecl

\bibitem{Mahdavinejad18MLI}
M.~S. Mahdavinejad, M.~Rezvan, M.~Barekatain, P.~Adibi, P.~M. Barnaghi, and
  A.~P. Sheth, ``Machine learning for internet of things data analysis: {A}
  survey,'' \emph{CoRR}, vol. abs/1802.06305, 2018.

\bibitem{Atzori10IoTS}
L.~Atzori, A.~Iera, and G.~Morabito, ``The internet of things: A survey,''
  \emph{Computer Networks}, vol.~54, no.~15, pp. 2787--2805, 10 2010.

\bibitem{Barile18Real-Time}
G.~Barile, A.~Leoni, L.~Pantoli, and V.~Stornelli, ``Real-time autonomous
  system for structural and environmental monitoring of dynamic events,''
  \emph{Electronics}, vol.~7, no. 12: 420, 2018.

\bibitem{Balducci18MLA}
F.~Balducci, D.~Impedovo, and G.~Pirlo, ``Machine learning applications on
  agricultural datasets for smart farm enhancement,'' \emph{Machines}, vol.~6,
  p.~38, 09 2018.

\bibitem{Brewster17IoT}
C.~Brewster, I.~Roussaki, N.~Kalatzis, K.~Doolin, and K.~Ellis, ``Iot in
  agriculture: Designing a europe-wide large-scale pilot,'' \emph{IEEE
  Communications Magazine}, vol.~55, no.~9, pp. 26--33, 2017.

\bibitem{Perkovic20Meeting}
T.~Perković, S.~Damjanović, P.~Šolić, L.~Patrono, and J.~Rodrigues,
  ``Meeting challenges in iot: Sensing, energy efficiency, and the
  implementation,'' in \emph{Fourth International Congress on Information and
  Communication Technology}.\hskip 1em plus 0.5em minus 0.4em\relax Springer,
  2020, pp. 419--430.

\bibitem{Dujic20MLS}
L.~D. Rodić, T.~Županović, T.~Perković, P.~Šolić, and J.~J. P.~C.
  Rodrigues, ``Machine learning and soil humidity sensing: Signal strength
  approach,'' (in press).

\bibitem{LiedmannHW18}
F.~Liedmann, C.~Holewa, and C.~Wietfeld, ``The radio field as a sensor - {A}
  segmentation based soil moisture sensing approach,'' in \emph{2018 {IEEE}
  Sensors Applications Symposium, {SAS} 2018, Seoul, South Korea, March 12-14,
  2018}.\hskip 1em plus 0.5em minus 0.4em\relax {IEEE}, 2018, pp. 1--6.

\bibitem{Mahdavinejad_ML_2018}
\BIBentryALTinterwordspacing
M.~S. Mahdavinejad, M.~Rezvan, M.~Barekatain, P.~Adibi, P.~Barnaghi, and A.~P.
  Sheth, ``Machine learning for internet of things data analysis: A survey,''
  \emph{Digital Communications and Networks}, vol.~4, no.~3, pp. 161 -- 175,
  2018. [Online]. Available:
  \url{http://www.sciencedirect.com/science/article/pii/S235286481730247X}
\BIBentrySTDinterwordspacing

\bibitem{kingma_adam_2014}
D.~Kingma and J.~Ba, ``Adam: A method for stochastic optimization,''
  \emph{International Conference on Learning Representations}, 2014.

\bibitem{glorot_bengio_2010}
X.~Glorot and Y.~Bengio, ``Understanding the difficulty of training deep
  feedforward neural networks,'' in \emph{Proceedings of the Thirteenth
  International Conference on Artificial Intelligence and Statistics}, ser.
  Proceedings of Machine Learning Research, vol.~9, 2010.

\bibitem{Hochreiter_lstm_1997}
S.~Hochreiter and J.~Schmidhuber, ``Long short-term memory,'' \emph{Neural
  Computation}, vol.~9, no.~8, pp. 1735--1780, 1997.

\bibitem{olah_lstm_2015}
\BIBentryALTinterwordspacing
C.~Olah. (2015) Understanding {LSTM} networks. [Online]. Available:
  \url{http://colah.github.io/posts/2015-08-Understanding-LSTMs/}
\BIBentrySTDinterwordspacing

\bibitem{tensorflow_2015}
\BIBentryALTinterwordspacing
M.~Abadi, A.~Agarwal, P.~Barham, E.~Brevdo, Z.~Chen, C.~Citro, G.~S. Corrado,
  A.~Davis, J.~Dean, M.~Devin, S.~Ghemawat, I.~Goodfellow, A.~Harp, G.~Irving,
  M.~Isard, Y.~Jia, R.~Jozefowicz, L.~Kaiser, M.~Kudlur, J.~Levenberg,
  D.~Man\'{e}, R.~Monga, S.~Moore, D.~Murray, C.~Olah, M.~Schuster, J.~Shlens,
  B.~Steiner, I.~Sutskever, K.~Talwar, P.~Tucker, V.~Vanhoucke, V.~Vasudevan,
  F.~Vi\'{e}gas, O.~Vinyals, P.~Warden, M.~Wattenberg, M.~Wicke, Y.~Yu, and
  X.~Zheng, ``{TensorFlow}: Large-scale machine learning on heterogeneous
  systems,'' 2015. [Online]. Available: \url{https://www.tensorflow.org/}
\BIBentrySTDinterwordspacing

\end{thebibliography}
\end{sloppypar}

\end{document}